\documentstyle[preprint,aps]{revtex}
\begin{document}
\title{On Exactness Of The Supersymmetric WKB Approximation Scheme}
\author{ R.S. Bhalla,
A.K. Kapoor \footnote {$^{}$ {email: akksp@uohyd.ernet.in}}
and P.K. Panigrahi \footnote {$^{}$ {email: panisp@uohyd.ernet.in}}}
\address {School of Physics,\\
University of Hyderabad,\\
Hyderabad - 500 046, India}
\maketitle
\begin{abstract}
Exactness of  the lowest order supersymmetric WKB (SWKB)
quantization condition $\int^{x_2}_{x_1} \sqrt{E-\omega^2(x)} dx
= n \hbar \pi$, for certain potentials, is examined, using
complex integration technique. Comparison of the above scheme
with a similar, but {\it exact} quantization condition, $\oint_c
p(x,E) dx = 2\pi n \hbar$, originating from the quantum
Hamilton-Jacobi formalism reveals that, the locations and the
residues of the poles that contribute to these integrals match
identically, for both of these cases. As these poles completely
determine the eigenvalues in these two cases, the exactness of
the SWKB for these potentials is accounted for. Three non-exact
cases are also analysed; the origin of this non-exactness is
shown to be due the presence of additional singularities in
$\sqrt{E-\omega^2(x)}$, like branch cuts in the $x-$plane.  \\

\vskip .5cm
\noindent
\flushleft
\vskip .25cm
\noindent
December 1995.
\end{abstract}
\newpage

Exactness of the lowest order supersymmetric WKB (SWKB) approximation
scheme for a class of potentials has been the subject of wide discussions
in the literature [1].
Explicit calculations, akin to that of the celebrated harmonic oscillator
case [2], first revealed the absence of O($\hbar^2 $) to O($\hbar^6$)
corrections to the lowest order result [3].
Subsequently, for one of the above potentials, it was shown that all higher
order corrections vanish identically [4].
A discrete reparameterisation symmetry, the so-called shape invariance, of
these potentials, has been thought to be responsible for this exactness
[5]. Recently, a new class of shape invariant potentials have been found,
for which the exactness of SWKB quantization does not hold true
[6]. In this light and considering the importance of this
quantization scheme in the potential problems, a deeper analysis
of the origin of this exactness is of utmost significance. \\

In supersymmetric (SUSY) quantum mechanics [7], a pair of
Hamiltonians, $H_\pm$ with potentials $V_\pm(x)$, given by
$\omega^2(x) \pm \hbar \partial \omega (x)/ \partial  x $,
($2m=1$), have identical bound state spectra, except for the
ground state;  $\omega (x)$ here, is the superpotential in SUSY
quantum mechanics. In the case of unbroken SUSY, conventionally,
the energy of the ground state of $H_-$, which is unpaired, is
taken to be zero. It was observed in [1], that a semi-classical
WKB type approximation $\int^{x_2}_{x_1} \sqrt{E-\omega^2(x)} dx
= n \hbar \pi$, making use of $ (V_-(x) + V_+(x))/2 =
\omega^2(x)$, gave  {\it exact} energy eigenvalues, for many well known
potentials. This is to be contrasted with the conventional WKB
scheme, which reproduced the exact spectra only for the harmonic
oscillator and the Morse potentials. \\

In what follows, we will first use a complex integration
technique, to evaluate the SWKB integral given by,
\begin{equation}
J_{SWKB} \equiv \frac{1}{\pi}\int_{x_1}^{x_2} \sqrt{E-\omega^2(x)}dx =
\frac{1}{2\pi}\oint_{C} \sqrt{E-\omega^2(x)}dx= n \hbar \, \, \, \, .
\label{1.9}
\end{equation}
\noindent
Here, $x_1$ and $x_2$ ($x_1<x_2$) are the turning points, which are the real
roots of $E-\omega^2(x) =0 $ and $C$ is a counter clockwise
contour, enclosing the branch cut between $x_1$ and $x_2$. Note
that the integrand is a double valued function, which is taken
to be positive just below the cut, joining the two turning
points [8].  The above integral will, then, be contrasted with
an exact quantization condition,
\begin{equation}
J(E) \equiv \frac{1}{2\pi}\oint_{C^\prime} p(x,E) dx = n \hbar \, \, \, \, ,
\label{1.9a}
\end{equation}
\noindent
originating from the quantum Hamilton-Jacobi (H-J) formalism
[9,10]. Here, $J(E)$ is the quantum action variable and $p(x,E)$
is the quantum momentum function (QMF) and $C^\prime$ is a
counter clockwise contour in the complex $x-$plane, enclosing
the real line  between the classical tuning points. The
classical turning points are the real values of $x$, for which
$E-V(x) \equiv p^2_c(x,E)$ vanishes, and encloses the region
between which the classical motion takes place. In a previous
paper [11] we used the quantization condition given by Eq.
($\ref {1.9a}$) to obtain the energy eigenvalues for a set of
SUSY potentials. This was done by rewriting the integral $J(E)$
in Eq. ($\ref {1.9a}$) as sum of integrals over contours
enclosing the singular points of QMF. \\

In this paper we evaluate $J_{SWKB}$ by complex integration
technique in a manner similar to that used in [11] for $J(E)$.
Interestingly, we find that the locations and the residues of
the poles of QMF, which determine the eigenspectra for the exact
quantization scheme match identically with the poles appearing
in the evaluation of the integral in Eq. ($\ref {1.9}$). This,
then, accounts for the SWKB scheme being exact in these cases.\\

In this paper  we also  analyse three potentials for which the
energy eigenvalues are known, but SWKB fails to give the exact
answer [5,12];  It is seen that the presence of additional
singularities, like branch cuts, prevents an exact calculation
of SWKB integrals for all these cases.  Intriguingly, in the
above examples, it is found that, if one ignores the
contributions of some of the branch cuts in the non-classical
region, the exact answer is reproduced. Hence, for these
potentials, the difference between the exact answer and the SWKB
approximation scheme, is quantified; this may be quite useful in
developing new approximation schemes for non-exactly solvable
problems. \\
The QMF satisfies the quantum H-J equation
$(2m=1)$,
\begin{eqnarray}
p^2(x,E) + \frac{\hbar}{i}\frac{\partial p(x,E)}{\partial x}
&=& E - V(x) \, \, \, \, \,  \nonumber \\
&\equiv& p^2_c(x,E)  \, \, \, \,  \, .
\label{1.3a}
\end{eqnarray}
\noindent
Here, $V(x)$ is any given potential and $p_c (x,E)$ is the
classical momentum function, defined so as to have the value
$+\sqrt{E-V(x)}$ just below the branch cut which
joins the two turning points. QMF $p(x,E)$ obeys
\begin{equation}
\lim_{\hbar \rightarrow 0} p(x,E) \rightarrow p_c(x,E) \, \, \, \, .
\label{1.3b}
\end{equation}
This can be thought of as a manifestation of the {\it
correspondence principle} or as a {\it boundary condition } on
the QMF. The quantum H-J equation is equivalent to the
Schr\"odinger equation, under the substitution,
\begin{equation}
p(x,E) = \frac \hbar i  \frac 1 {\psi (x,E)} \frac {\partial \psi
(x,E)} {\partial x} \, \, \, \, .
\end{equation}
The nodes of the wave function $\psi(x,E)$ are known to appear between the
classical turning points. QMF $p(x,E)$ will then have poles at the
corresponding points, hence  $J(E)$ which counts this number, will
obey the exact quantization condition, given in Eq. ($\ref
{1.9a}$). If we substitute an $\hbar$ expansion for $p(x,E)$,
the quantization condition ($\ref {1.9a}$) coincides with the
well known Dunham's formula. \\

We explicitly workout the SWKB integral in Eq. ($\ref {1.9}$)
for a set of trigonometric and hyperbolic potentials and
tabulate the results in Table I. In what follows, the steps
required to solve one of these potentials are  outlined. We then
compare these results with our previous calculation [11] using the
quantum action variable.  \\

We shall now evaluate $J_{SWKB}$ for the Eckart potential,
\begin{equation}
V_-(x) = A^2 + \frac {B^2} {A^2} +A(A-\alpha \hbar)
{\rm cosech}^2 \alpha x - 2 B \coth \alpha x \, \, \, \, .
\end{equation}
whose superpotential is,
\begin{equation}
\omega (x) = - A \coth \alpha x + B/A \, \, \, \, \, ,
\hspace {1.cm}  B>A^2  \, \, \, {\rm and} \, \, \,
0<x< \infty  \, \, \, \, .
\label{1.10}
\end{equation}

We use the mapping $y = \exp(\alpha x)$ in which case,

\begin{equation}
V(y) = A^2 + \frac{B^2}{A^2}
+\frac{4A(A- \alpha \hbar)y^2}{(y^2-1)^2}
- \frac{2B(y^2+1)}{(y^2-1)} \, \, \, \, \, \, \, ,
\end{equation}
and
\begin{equation}
\omega (y) = - A \frac{y^2+1}{y^2-1} +\frac{B}{A} \, \, \, \, .
\label{1.12}
\end{equation}
The quantization condition in Eq. ($\ref{1.9}$) for the SWKB
approximation becomes,
\begin{equation}
J_{SWKB}  = \frac{1}{2\pi\alpha}\oint_{C_1} \frac{\sqrt{E-\omega^2(y)}}{y}dy
               = n \hbar \,\, \, \, .
\label{1.15}
\end{equation}
\noindent

\noindent
It is clear from Eq. ($\ref {1.15}$) that, the integrand has
singularities at $y=0$ and $y= \pm 1$. The roots of
$E-\omega^2(y) = 0$, give the branch points; they are four in
number, two of which lie in the classical and the other two in
the non-classical $(y<0)$ region of the complex $y-$plane. To evaluate
the above integral, consider a counter clockwise contour
integral $J_{\Gamma_R}$ for a circle $\Gamma_R$ of radius $R$,
which is such that it encloses all the singular points of the
above integrand (see Fig. 1). Then,
\begin{equation}
J_{\Gamma_R} = J_{SWKB} + J_{C_2} + J_{\gamma_1} + J_{\gamma_2} +
J_{\gamma_3}
\, \, \, \, \, .\label{1.17}
\end{equation}
\noindent
Here, $J_{C_2}$ is the integral along
the counter clockwise contour $C_2$ enclosing the branch cut in
the Re$y <0$ region. $ J_{\gamma_1}, J_{\gamma_2}$ and $J_{\gamma_3}$
are the integrals along contours $ \gamma_1, \gamma_2$ and
$\gamma_3$ enclosing the singular points at $y=1$, $y=-1$ and
$y=0$ respectively. It may be noticed that the symmetry $y
\rightarrow - y$ of $E-\omega^2(y) $ implies,
\begin{equation}
J_{SWKB} = J_{C_2}  \, \, \, \, .
\label{1.16}
\end{equation}
For the pole at $y=0$, one gets,
\begin{equation}
J_{\gamma_1} = \frac{i \sqrt{E - A^2 - B^2/A^2 - 2B}}{\alpha} \, \,
\, \, .
\label{1.22}
\end{equation}
\noindent
Similarly for poles at $y = \pm 1$,
\begin{equation}
J_{\gamma_2} = J_{\gamma_3} = \frac {A}{\alpha} \, \, \, \, \, .
\label{1.20}
\end{equation}
\noindent
Now for the calculation of $J_{\Gamma_R}$ one more change of
variable $z = 1/y$ is sought; so the singularity at $y
\rightarrow \infty$ is mapped to the singularity at $z=0$. The
contour integral for the pole at $z=0$ is,
\begin{equation}
J_{\Gamma_R} = \frac{-i \sqrt{E - A^2 - B^2/A^2 + 2B}}
{\alpha} \, \, \, \, \, .
\label{1.23}
\end{equation}
\noindent
The SWKB quantization condition
\begin{equation}
J_{\Gamma_R} - \sum_{p=1}^{3} J_{\gamma_p} = 2 n \hbar \, \, \, \, \, ,
\label{1.24}
\end{equation}
\noindent
when inverted for $E$ gives
\begin{equation}
E_n = A^2 + B^2/A^2 - \frac{B^2}{(n \alpha \hbar + A)^2} -
(n \alpha \hbar + A)^2 \, \, \, \, .
\label{1.25}
\end{equation}
\noindent
The calculation of eigenvalues for other SUSY potentials
proceeds along similar lines and the results are summarised in
Table I. \\
Comparing the above computation of $J_{SWKB}$, with that of
$J(E)$ in our previous paper, we find that the singularity
structure of $\sqrt{E-\omega^2}$ other than the branch cuts,
match exactly with that of the fixed poles of $p(x,E)$ in the
quantum H-J formalism [11]. In all the cases tabulated in Table
I, the location of the fixed poles and the values of the
corresponding residues are identical for both the SWKB
approximation and the quantum H-J formalism. The contour
integral $J(E)$,  matches exactly with the $J_{SWKB}$. This then
explains the rather mysterious result of the exactness of SWKB
approximation for the potentials in Table I. \\

We next consider the cases where lowest order SWKB fails to give
exact answer. We at first look at the potential [12]
\begin{equation}
V(x) = \frac{1}{4} (x^2 + \frac{3}{4x^2}) - 2 + \frac{(3x^2
+x^4)}{(1+x^4)} \, \, \, \, ,
\label{1.26}
\end{equation}
with
\begin{equation}
\omega (x) = \frac{2x^6 + 3x^4 -x^2 - 6}{2x(x^2 +1)(x^2 + 2)} \,
\, \, \, \, .
\label{1.27}
\end{equation}
\noindent
The SWKB quantization condition reads
\begin{equation}
J_{SWKB} = \frac{1}{2 \pi} \oint_C \sqrt{E-\omega^2(x)} dx = n
\hbar \, \, \, \, \, \, ,
\label{1.28}
\end{equation}
\noindent
It is noticed that the above integrand becomes zero at twelve
points in the complex plane giving rise to six branch cuts. One
of these branch cuts is included in the contour $C$. The integral
around a branch cut on the negative axis is equal to the
above integral due  to the symmetry $x \rightarrow -x$.  \\
\noindent
The integrals around the other four branch cuts cannot be
related to the integral $J_C $. We use $J_{OBC}$ to denote
the sum of the integrals along contours enclosing the four
remaining branch cuts. \\

We now try to proceed, as far as possible, parallel to the Eckart
potential. Let $J_ {\Gamma_R}$ be a circular contour with the center
at origin and radius large enough to enclose all the singular
points of the integrand. Note that the integrand has poles at
$y=0$; $y=\pm i$ and $y =\pm
\sqrt{2}i$. The contour integral $J_{\Gamma_R}$ is given by
\begin{equation}
J_{\Gamma_R} = 2n \hbar + J_{OBC} + J_{\gamma_1} + J_{\gamma_2}
+ J_{\gamma_3}   + J_{\gamma_4} + J_{\gamma_5}  \, \, \, \, ,
\label{1.32}
\end{equation}
or
\begin{equation}
J_{\Gamma_R} - \sum_{p=1}^5 J_{\gamma_p} = 2 n \hbar - J_{OBC}
\, \, \, \, ,
\label{1.32b}
\end{equation}
where   $J_{\gamma_1}, J_{\gamma_2}, J_{\gamma_3}, J_{\gamma_4}$ and
$J_{\gamma_5}$ are the corresponding contour integrals for the
contours $\gamma_1, \gamma_2, \gamma_3, \gamma_4$ and
$\gamma_5$, around $y=0; +i ; -i ;+\sqrt{2} i$ and - $\sqrt{2}i$  \\

\noindent
Various contour integrals can be evaluated as before and we get
$J_{\gamma_1} = 3/2 \, \, , J_{\gamma_2} = -1  \, \, ,
J_{\gamma_3} = -1   \, \, , J_{\gamma_4} = 1   \, \, $ and $
J_{\gamma_5} = 1 $.  The signs of the residues needed have been
fixed as for the Eckart potential. $J_{\Gamma_R}$ is again
calculated making the mapping $y \rightarrow 1/z $ and
looking for the residue of the pole at $z \rightarrow 0$. The
corresponding contour integral is found to be,
\begin{equation}
J_{\Gamma_R} = 2E + 3/2 \, \, \, \, .
\label{1.31}
\end{equation}
The SWKB quantization condition is equivalent to Eq. ( $\ref {1.32b}$).
and this condition involves the unknown integral $J_{OBC}$.
Therefore, this complex variables approach cannot be used to
complete the calculation of energy eigenvalues in SWKB approximation for
the potential ($\ref {1.26}$).  However, it is easily checked that if we
substitute the values of $J_{\Gamma_R}, J_{\gamma_1},
J_{\gamma_2}, J_{\gamma_3}, J_{\gamma_4}$ and $J_{\gamma_5}$ in
\begin{equation}
J_{\Gamma_R} - \sum_{p=1}^5 J_{\gamma_p} = 2 n \hbar
\, \, \, \, ,
\label{1.33}
\end{equation}
\noindent
we obtain the exact eigenvalues. \\
Therefore, it is clear that if the
sum of the contributions coming from other branch cuts is nonzero,
the SWKB approximation will not be exact. \\

\noindent
A similar analysis can be repeated for some of the other
potentials. It is known that [12,5] for
\begin{equation}
V(x) = -1/x + \frac{x(x+2)} {(1+x + x^2/2 )^2} +1/16 \, \, \, \, ,
\label{1.34}
\end{equation}
and
\begin{equation}
V(x) = (1-y^2)\left[-\lambda^2 \nu (\nu +1) + \frac{1}{4}(1-\lambda^2)
[2 -(7-\lambda^2)y^2 +5(1-\lambda^2)y^4] \right] \, \, \, \, ,
\label{1.38}
\end{equation}
with super potentials,
\begin{equation}
\omega (x) = \frac{x^6 -16 x^4 -56 x^3 -108 x^2 -240 x -192}
{4x(x^2+2x+2)(x^3+6x^2+16x+24)} \, \, \, \, ,
\label{1.35}
\end{equation}
and
\begin{equation}
\omega (x) = \frac{1}{2}(1-\lambda^2)y(y^2-1) \mu_0 \lambda^2 y
\, \, \,  \,  ,
\label{1.40}
\end{equation}
SWKB does not give exact answers. We once again see that SWKB is
not exact because of the contributions from the other branch cuts,
that are present in both these cases.\\
\newpage
\noindent
{\bf V. CONCLUSION: }  \\
\noindent
To conclude, we have shown using the complex contour integration
technique, for several solvable potentials in one dimension,
that SWKB gave exact results. We then compared it with the
answers gotten using the quantum H-J method, from the {\it
exact} quantization condition.  This has provided some useful
insight: It is the similarity of the singularity structure of
$p(x,E)$ and $\sqrt{E-\omega^2 }$ in the non-classical
regions of the $x-$plane, namely the matching of the poles and
the residues that is responsible for this exactness. Table I
contains all the relevant details about the steps involved in
the calculation.\\ For the non-exact potentials, it is the
inability to deform the contour appropriately because of the
presence of other poles and branch cuts in the complex $x-$plane
that prevents an exact solution in SWKB approach.  \\ {\bf
Acknowledgements:} We acknowledge useful discussions with Prof.
V.  Srinivasan and Dr. C.N. Kumar.  \\

\noindent{\bf References}
\begin{enumerate}
\item A. Comtet, A. Bandrauk and D. Campbell, Phys. Lett. {\bf
B150}, 159 (1985); \\ A. Khare, Phys. Lett. {\bf B161}, 131 (1985).
\item C.M. Bender, K. Olaussen and P.S. Wang, Phys. Rev. {\bf
D6}, 1740 (1977).
\item R. Adhikari, R. Dutt and U.P. Sukhatme, Phys. Rev. {\bf
A38} 1679 (1988).
\item K. Raghunathan, M. Seetharaman and S.S. Vasan, Phys. Lett.
{\bf B188}, 351 (1987).
\item A. Khare and Y.P. Varshni, Phys. Lett. {\bf A142}, 1 (1989).
\item D.T. Barclay, A. Khare and U. Sukhatme, Phys. Lett {\bf
A183}, 263 (1993).
\item E. Witten, Nucl. Phys. {\bf B188}, 513 (1981). ; \\
F. Cooper, A. Khare and U. Sukhatme, Phys. Rep. {\bf 251} 267 (1995).
and references contained therein.
\item See, for example,  H. Goldstein, {\it Classical
Mechanics} (Addison Wesley, New York, 1980) p. 474.
\item R.A. Leacock and M.J. Padgett Phys. Rev. Lett.  {\bf 50}, 3 (1983);
R.A. Leacock and M.J. Padgett Phys. Rev.  {\bf D28}, 2491 (1983).
\item For a pedagogical  review  see, R.S. Bhalla, A.K. Kapoor
and P.K. Panigrahi, University of Hyderabad preprint, Dec. 1995.
\item R.S. Bhalla, A.K. Kapoor and P.K. Panigrahi, hep-th/9507154, 1995.
\item David De Laney and M.M. Nieto. Phys. Lett. {\bf B247} 301 (1990).

\end{enumerate}
\pagebreak

\pagestyle{empty}
\leftmargin=-1truein
\textwidth=7.5truein
\textheight=10truein
\flushleft
\setlength{\oddsidemargin}{-0.25 in}
\setlength{\evensidemargin}{1.0 in}
\baselineskip=12pt
Table I :  Hyperbolic and trigonometric potentials.
The mapping used for hyperbolic potentials is $y=\exp(\alpha x)$, while
for the trigonometric ones $y=\exp(i \alpha x)$ is being used; \, \, \, \, \,
\, \, \,
$\alpha$ is real and  positive.
\begin{tabular}{ l c c c c }
\hline \hline
Name of & Super potential & Fixed
& $\alpha I_{\gamma_p}$ & Eigenvalue \\
potential &  &  poles at
&  & $E_n$ \\
\hline
 &  &   0 & $ -i \sqrt{(E-A^2)}$ &    \\
Scarf II  &  &  $i$ &
$ (iB -A)$ &   \\
(hyperbolic) &
$ A \tanh \, \alpha x + B {\rm sech} \, \alpha x$
&$-i$ &$-(iB+A)$ & $ A^2 -(A -\alpha
\hbar)^2 $   \\
 &  & $ \infty $   & $ i \sqrt{(E-A^2)} $ &     \\
 & & &  & \\
Rosen - &  & $0$ & $ - i \sqrt{E-(A
-\frac{B}{A})^2 }$ &
$ A^2 + B^2 /A^2  $ \\
  Morse II & $A \tanh \, \alpha x + B/A $ & $i$ & $-A$ & $ - (A -
\alpha \hbar )^2 $ \\
(hyperbolic) & $(B \, < \, A^2) $ &  $-i$   & $ - A $ & $ - B^2/(A- n\alpha
\hbar)^2$   \\
 &  &  $\infty $ & $ i \sqrt{E-(A +\frac{B}{A})^2 } $ &   \\
 & & & & \\
Generalised &  &  $0$   & $ - i \sqrt{(E-A^2)}$ &   \\
Poschl-& $ A \coth \, \alpha x - B{\rm cosech}^2 \,\alpha x $
&$1$& $-(A-B)$ & $ A^2 -(A -  n\alpha
\hbar )^2  $ \\
Teller & $ (A\, < \, B  \, \, {\rm and} \, \, x \, \geq \, 0) $
   & $-1$   & $ -(A+B) $ &  \\
(hyperbolic)& & $\infty$ & $ i \sqrt{(E-A^2)} $ &\\
 & & & & \\
Scarf I  &  &   $0$   & $   \sqrt{(E+A^2)}$ &  \\
(trig-& $-A \tan \, \alpha x + B \sec \, \alpha x $ & $i$  &$-(A-B) $ &
$ (A + n\alpha
\hbar )^2 - A^2 $ \\
onometric) & ($- \pi/2 \le \alpha x \le \pi /2 $) & $-i$
& $ -(A + B) $ &
\\ &   &$ \infty $ & $
-\sqrt{(E+A^2)}$ & \\
 & & & & \\
Rosen- &  &  $0$   & $ - \sqrt{E+(A +i\frac{B}{A})^2}$ &  $
A^2 + B^2 /A^2   $\\
Morse-I & $- A \cot \, \alpha x  -B/A  $ &$1$ & $A$ &
$  -(A +n\alpha \hbar )^2 $ \\
 (trig- & ($0 \le \alpha x \le \pi$) &  $-1$ & $ A $ & $  -B^2/(A+ n\alpha
\hbar )^2$  \\
onometric)&  &
 $\infty $ & $  \sqrt{E+(A -i\frac{B}{A})^2} $ & \\
\hline
\hline
\end{tabular}
\samepage
\pagebreak

{\bf Figure Captions} \\
\begin{enumerate}
\item  Fig.1. Contour for the Eckart potential problem.

\end{enumerate}
\end{document}